\documentclass{PoS}

\title{Algebraic geometry informs perturbative quantum field theory}

\ShortTitle{Algebraic geometry informs perturbative quantum field theory}

\author{\speaker{David Broadhurst}\\
Open University, UK, and Humboldt-Universit\"at zu Berlin\\
E-mail: \email{David.Broadhurst@open.ac.uk}}

\author{Oliver Schnetz\\
University of Erlangen and Humboldt-Universit\"at zu Berlin\\
E-mail: \email{schnetz@mi.uni-erlangen.de}}

\abstract{Single-scale Feynman diagrams yield integrals that
are periods, namely projective integrals of rational functions
of Schwinger parameters. Algebraic geometry may therefore inform
us of the types of number to which these integrals evaluate. We give examples
at 3, 4 and 6 loops of massive Feynman diagrams that evaluate
to Dirichlet $L$-series of modular forms and examples at 6, 7 and 8 loops
of counterterms that evaluate to multiple zeta values or polylogarithms
of the sixth root of unity. At 8 loops and beyond, algebraic geometry informs
us that polylogs are insufficient for the evaluation of terms in the
beta-function of $\phi^4$ theory. Here, modular forms appear as obstructions
to polylogarithmic evaluation.}

\dedicated{In memoriam Jochem Fleischer (1937-2013)}

\FullConference{Loops and Legs in Quantum Field Theory\\
27 April 2014 - 02 May 2014\\
Weimar, Germany}

\begin{document}

\newcommand{\et}[2]{\eta_{#1}^{#2}}
\newcommand{\Df}[2]{\mbox{$\frac{#1}{#2}$}}

\section{Beyond polylogs: modular forms}

We begin with a short guide to modular forms.
For  $|q|<1$, let
$$\eta(q)\equiv q^{1/24}\prod_{n>0}(1-q^n)=\sum_{n=-\infty}^\infty(-1)^n q^{(6n+1)^2/24}$$
then for $\Im z>0$, 
$$\eta(\exp(2\pi{\rm i}z))=({\rm i}/z)^{1/2}\eta(\exp(-2\pi{\rm i}/z)).$$
If $f(z)=(\sqrt{-N}/z)^w f(-N/z)$, we say that $f$ is a modular
form of modular weight $w$ and level $N$. Here is a well known example with
modular weight 12 and level 1:
$$[\eta(q)]^{24}=\sum_{n>0}A(n)q^n=
q - 24q^2 + 252q^3 - 1472q^4 + 4830q^5 - 6048q^6 - 16744q^7+\ldots$$
Its Fourier coefficients  are multiplicative:
$A(m n)=A(m)A(n)$ for ${\rm gcd}(m,n)=1$,
and are determined by $A(p)$ at the primes $p$: 
$$L(s)\equiv \sum_{n>0}\frac{A(n)}{n^s}
=\prod_{p}\frac{1}{1-A(p)p^{-s}+p^{11-2s}}.$$ 
Moreover, we  can analytically continue to values inside the critical strip: 
$$\Lambda(s)\equiv \frac{\Gamma(s)}{(2\pi)^s}L(s)
=\sum_{n>0}A(n)\int_1^\infty{\rm d}x\left(x^{s-1}+x^{11-s}\right)\exp(-2\pi n x)=\Lambda(12-s).$$

\subsection{Multiplicative modular forms from eta-products}

For brevity, let $\eta_n\equiv\eta(q^n).$
Here are some multiplicative modular forms identified in
quantum field theory
$$\begin{array}{lccl}
{\rm form}&{\rm weight}&{\rm level}&{\rm QFT}\\
\et13\et73&3&7&{\rm BS}\\
\et12\et2{}\et4{}\et82&3&8&{\rm BS}\\
\et23\et63&3&12&{\rm BS}+{\rm BFT}+{\rm BBBG}+{\rm BV}\\
\et14\et54&4&5&{\rm BS}\\
\et12\et22\et32\et62&4&6&{\rm BS}+{\rm BB}\\
\et14\et22\et44&5&4&{\rm BS}\\
\et16\et36&6&3&{\rm BS}\\
\et2{12}&6&4&{\rm BS}\\
\et18\et28&8&2&{\rm BS}\\
\et1{24}&12&1&{\rm BK}
\end{array}$$
from work by
Bailey, Borwein, Broadhurst, Glasser~\cite{BBBG} (BBBG),
Bloch, Vanhove~\cite{BV} (BV),
Broadhurst, Brown~\cite{B3} (BB),
Broadhurst, Fleischer, Tarasov~\cite{BFT} (BFT), 
Broadhurst, Kreimer~\cite{BK1,BK2} (BK),
Brown, Schnetz~\cite{BS1,BS2} (BS).

{\bf Remark:} QFT seems blind to Birch and Swinnerton--Dyer: nothing appears at weight 2.

\section{Beyond polylogs and elliptic integrals} 

Consider the two-loop massive sunrise diagram in $D=2$ spacetime dimensions:
$$I(p^2,m_1,m_2,m_3)\equiv \frac{1}{\pi^2}\left(\prod_{k=1}^3\int
\frac{{\rm d}^2q_k}{q_k^2-m_k^2+{\rm i}\epsilon}\right)
\delta^{(2)}(p-q_1-q_2-q_3).$$
Following BBBG, we obtain an efficient result  from the discontinuity across the cut~\cite{BBBG}:
$$I(w^2,m_1,m_2,m_3)=8\pi\int_{m_1+m_2+m_3}^\infty
\frac{A(x)x{\rm d}x}{x^2-w^2}\label{CG}$$
with an elliptic integral yielding the reciprocal 
$$A(w)=\frac{2}{\pi}\int_0^{\pi/2}\frac{{\rm d}\theta}{\sqrt{F(w)\cos^2\theta
+16m_1m_2m_3w \sin^2\theta}}=\frac{1}{{\rm agm}\left(\sqrt{F(w)},\sqrt{F(w)-F(-w)}\right)}$$ 
of an \emph{arithmetic-geometric mean} with
$$F(w)=(w+m_1+m_2+m_3)(w+m_1-m_2-m_3)(w-m_1+m_2-m_3)(w-m_1-m_2+m_3).$$
From the complementary elliptic integral 
$$B(w)=\frac{1}{{\rm agm}\left(\sqrt{F(w)},\sqrt{F(-w)}\right)}$$
we obtain the \emph{elliptic nome} 
$$q(w)\equiv\exp(-\pi B(w)/A(w)).$$

\subsection{Differential equation in the equal--mass case}

Now set $m_1=m_2=m_3=1$.  Then $F(w)=(w+3)(w-1)^3$
and the differential equation, found in 1993 with 
Jochem Fleischer (sadly deceased on 2 April 2013) and Oleg Tarasov~\cite{BFT}, gives 
$$-\left(\frac{q(w)}{q^\prime(w)}\frac{\rm d}{{\rm d}w}\right)^2\left(\frac{I(w^2,1,1,1)}{24\sqrt{3}A(w)}\right)
=\frac{w^2(w^2-1)(w^2-9)A(w)^3}{9\sqrt{3}}.$$
Regarding $w$ and $A(w)$ as functions of $q$, 
we have a parametric solution 
$$\frac{w}{3}=\left(\frac{\eta_3}{\eta_1}\right)^4\left(\frac{\eta_2}{\eta_6}\right)^2,\quad
4\sqrt{3}A=\frac{\et16\et6{}}{\et23\et32}.$$
Moreover, the two algebraic relations between $\{\eta_1,\eta_2,\eta_3,\eta_6\}$ give
$$\frac{w^2-1}{8}=\left(\frac{\eta_2}{\eta_1}\right)^9\left(\frac{\eta_3}{\eta_6}\right)^3,\quad
\frac{w^2-9}{72}=\left(\frac{\eta_6}{\eta_1}\right)^5\frac{\eta_2}{\eta_3}.$$
Hence the BFT differential equation reduces to
$$-\left(q\frac{{\rm d}}{{\rm d}q}\right)^2\left(\frac{I}{24\sqrt{3}A}\right)=\frac{w}{3}f_{3,12}
=\left(\frac{\et33}{\eta_1}\right)^3+\left(\frac{\et63}{\eta_2}\right)^3$$
where, remarkably, $f_{3,12}\equiv(\eta_2\eta_6)^3$ is a weight-3 level-12 modular form found in 
massless $\phi^4$ theory by Brown and Schnetz at 9 loops~\cite{BS2}.

\subsection{Bloch--Vanhove elliptic dilogarithm}

Define a character with $\chi(n)=\pm1$ for $n=\pm1$ mod 6 and $\chi(n)=0$ otherwise. Then
$$-\left(q\frac{{\rm d}}{{\rm d}q}\right)^2\left(\frac{I}{24\sqrt{3}A}\right)=
\sum_{n>0}\frac{n^2(q^n-q^{5n})}{1-q^{6n}}=\sum_{n>0}\sum_{k>0}n^2\chi(k)q^{nk}.$$
Integrating twice and using the known imaginary part on the cut, we recover the BV result~\cite{BV}
$$\frac{I(w^2,1,1,1)}{4A(w)}=E_2(q)=-\pi\log(-q)-3\sqrt{3}\sum_{k>0}\frac{\chi(k)}{k^2}\frac{1+q^k}{1-q^k}=-E_2(1/q)$$
with constants of integration that make $I$ finite at the pseudo-threshold, where $q=-1$.

\subsection{An elliptic trilogarithm at 3 loops}

The equal-mass three-loop sunrise integral $J(t)$ yields an elliptic trilogarithm for~\cite{BKV}
$$\frac{2J(t)}{\omega_1(t)}=E_3(q)=(-2\log(q))^3+\sum_{k>0}\frac{\psi(k)}{k^3}\frac{1+q^k}{1-q^k}=-E_3(1/q)$$
with $\psi(k)=\psi(k+6)=\psi(-k)$, 
$\psi(1)=-48$, $\psi(2)=720$, $\psi(3)=384$, $\psi(6)=-5760$ and
$$q=\exp\left(-\frac{2\pi}{3}\frac{A(\widetilde{w})}{B(\widetilde{w})}\right)$$ where $A$ and $B$ are the elliptic integrals
for the equal-mass two-loop case, but now evaluated at 
$$\widetilde{w}=\sqrt{1-t/4}+\sqrt{4-t/4}.$$
Then the transformation between the Green functions
for hexagonal and diamond lattices given by BBBG in Eq.~(188) of~\cite{BBBG} provides 
$\omega_1(t)=(\widetilde{w}B(\widetilde{w}))^2$ as a solution to the homogeneous 
equation that is regular as $t\to\infty$. Underlying the expression for $q$
are the cubic and sesquiplicate modular transformations of~\cite{B2}. The constants
of integration for $E_3(q)$ are determined by the requirement that $J(t)$ is finite
as $t\to0$ and reproduces the value $J(0)=7\zeta(3)$ proven by BBBG~\cite{BBBG}.

\section{Modular forms in massive QFT}

Here we consider multi-loop on-shell sunrise diagrams in two spacetime dimensions. 
At $L$ loops, these are given
in coordinate space as single integrals over $N=L+2$ Bessel functions: $L+1$ copies of
$K_0(x)$ from the internal lines and a single $I_0(x)$ from Fourier transformation with 
respect the external on-shell momentum. More generally, let 
$$S_{N,L}\equiv2^L\int_0^\infty I_0(x)^{N-L-1}K_0(x)^{L+1}x{\rm d}x.$$
Then $S_{6,4}$ is indicative of some of the number theory
entering $g-2$ at 4 loops, where Stefano Laporta is tackling
diagrams with 5 fermions in the intermediate state.

For convergence, we require that  $L<N\le2L+2$. With $N=2L+2$
we require that $L>1$.  BBBG proved that~\cite{BBBG}
$$S_{1,0}=S_{2,1}=1,\quad
S_{3,1}=\frac{2\pi}{3\sqrt{3}},\quad
S_{3,2}=\frac{4\,{\rm Cl}_2(\pi/3)}{\sqrt{3}},\quad
S_{4,2}=\frac{\pi^2}{4},\quad
S_{4,3}=7\zeta(3),$$
$$S_{5,2}=\frac{\pi^2}{8}\left(\sqrt{15}-\sqrt{3}\right)
\left(\sum_{n=-\infty}^\infty{\rm e}^{-\sqrt{15}\pi n^2}\right)^4
=\frac{\sqrt3}{120\pi}\
\Gamma(1/15)\Gamma(2/15)\Gamma(4/15)\Gamma(8/15)$$
where the final product of Gamma values results from the
Chowla--Selberg theorem. York Schr\"oder needed
a counterterm in 3-dimensional lattice field theory\cite{DLST}
for which we used Chowla--Selberg to obtain
the product $(\Gamma(1/24)\Gamma(11/24))^2$.

BBBG also conjectured (and checked to 1000 digits) that~\cite{BBBG}
$$S_{5,3}=\frac{4\pi}{\sqrt{15}}S_{5,2},\quad
S_{6,4}=\frac{4\pi^2}{3}S_{6,2},\quad
S_{8,5}=\frac{18\pi^2}{7}S_{8,3}.$$

{\bf Remark:} In 4-dimensions, there are two master on-shell 3-loop sunrise integrals
Amazingly, the finite parts
of their Laurent expansions are linear combinations of
$S_{5,3}$ and $\pi^6/S_{5,3}$, with the reciprocal coming from
$\Gamma(7/15)\Gamma(11/15)\Gamma(13/15)\Gamma(14/15)$.

\subsection{Sunrise at 3 loops from a modular form of weight 3}

Let $L_{3,15}(s)$ be the Dirichlet $L$-function defined by the 
multiplicative modular form
$$f_{3,15}=(\eta_3\eta_5)^3+(\eta_1\eta_{15})^3$$
with weight 3 and level 15.  
Then we conjecture (and have checked to 1000 digits) that~\cite{B3}
$$S_{5,2}=3L_{3,15}(2),\quad S_{5,3}=\frac{8\pi^2}{15}L_{3,15}(1),$$
where $S_{5,3}$ is the 5-Bessel moment giving  the on--shell 3--loop sunrise diagram.

\subsection{Sunrise at 4 loops from a modular form of weight 4}

Let $L_{4,6}(s)$ be the Dirichlet $L$-function defined by the multiplicative modular form
$$f_{4,6}=(\eta_1\eta_2\eta_3\eta_6)^2$$
with weight 4 and level 6.  
Then we conjecture (and have checked to 1000 digits) that~\cite{B3}
$$S_{6,2}=6L_{4,6}(2),\quad S_{6,3}=12L_{4,6}(3),\quad
S_{6,4}=8\pi^2L_{4,6}(2),$$
where $S_{6,4}$ is the 6-Bessel moment giving  the on--shell 4--loop sunrise diagram.

\subsection{Almost sunrise at 6 loops from a modular form of weight 6}

Let $L_{6,6}(s)$ be the Dirichlet $L$-function defined by the multiplicative modular form
$$f_{6,6}=
\left(\frac{\et23\et33}{\eta_1\eta_6}\right)^3+
\left(\frac{\et13\et63}{\eta_2\eta_3}\right)^3$$
with weight 6 and level 6.  
Then we conjecture (and have checked to 1000 digits) that~\cite{B3}
$$S_{8,3}=8L_{6,6}(3),\quad S_{8,4}=36L_{4,6}(4),\quad
S_{8,5}=216L_{4,6}(5),$$
but have no result for $S_{8,6}$, the 8-Bessel moment for the on--shell 6--loop sunrise diagram.

\section{Counterterms at 6 to 10 loops }

\subsection{6 loops: the first MZV}

In 1995, Broadhurst and Kreimer evaluated all the counterterms for the coupling of
$\phi^4$ theory coming from subdivergence-free diagrams up to
6 loops~\cite{BK1}. All were expressible in terms
of $\zeta(3)$, $\zeta(5)$, $\zeta(7)$, $\zeta(9)$ and $\zeta(5,3)-\Df{29}{12}\zeta(8)$,
where the multiple zeta value $\zeta(5,3)=\sum_{m>n>0}m^{-5}n^{-3}$
is not reducible to single zeta values and occurs at 6 loops.

\subsection{7 loops: unexpected products}

At 7 loops, $\zeta(7,3)-\Df{793}{94}\zeta(10)$, $\zeta(11)$ and  $\zeta(3,5,3)-\zeta(3)\zeta(5,3)$ appear~\cite{BK1,BK2}.
In 1995, three 7-loop counterterms were lacking:
$P_{7,8}$,  $P_{7,9}$ and $P_{7,11}$ in the Schnetz census~\cite{S1}. Allowing for a product of $\zeta(3)$
with a 6-loop counterterm,  we eventually determined
\begin{eqnarray*}
P_{7,8}&=&\frac{22383}{20}\zeta(11)+\frac{4572}{5}\left[\zeta(3,5,3)-\zeta(3)\zeta(5,3)\right]-700\zeta(3)^2\zeta(5)\nonumber\\
&&\quad+\,1792\zeta(3)\left[\frac{9}{320}\left[12\zeta(5,3)-29\zeta(8)\right]+\frac{45}{64}\zeta(5)\zeta(3)\right],\\
P_{7,9}&=&\frac{92943}{160}\zeta(11)+\frac{3381}{20}\left[\zeta(3,5,3)-\zeta(3)\zeta(5,3)\right]-\frac{1155}{4}\zeta(3)^2\zeta(5)\nonumber\\
&&\quad+\,896\zeta(3)\left[\frac{9}{320}\left[12\zeta(5,3)-29\zeta(8)\right]+\frac{45}{64}\zeta(5)\zeta(3)\right].
\end{eqnarray*}

\subsection{7 loops:  failures of the Kontsevich conjecture}

A sub-divergence free $\phi^4$ counterterm is of interest to algebraic geometers
who regard it as a rather special type  of ``period". At 7 loops, with 14 edges,
we have an integral over 13 Schwinger parameters:
$$P_G=\int_{\alpha_i>0}\frac{{\rm d}\alpha_2\ldots{\rm d}\alpha_{14}}
{\left|\Psi_G(\alpha_1,\alpha_2,\ldots,\alpha_{14})\right|_{\alpha_1=1}^2}$$
where physicists call $\Psi_G$ the first Symanzik polynomial and graph theorists call it
the Kirchhoff polynomial of the graph $G$. Suppose that
we  count the number $c(q)$ of solutions to $\Psi_G=0$ 
in finite fields ${\bf F}_q$ for prime powers  $q=p^k$. 
If $c(q)$ is a polynomial in $q$, then algebraic
geometers expect the period $P_G$ to evaluate to MZVs. 

Maxim Kontsevich conjectured that $c(q)$ is a polynomial
for every graph $G$.  This was proved true by Stembridge, by exhaustion,
for all graphs with no more than 12 edges~\cite{St}. It is true for all 7-loop periods
except~\cite{S2} $P_{7,8}$,  $P_{7,9}$ and $P_{7,11}$.

It is the prime $p=2$ that makes the periods $P_{7,8}$ and $P_{7,9}$
non-Stembridge~\cite{D,S2}. Thus they were expected to be reducible to alternating
sums in the MZV datamine~\cite{BBV}. Algorithms due to Panzer and Schnetz
have achieved this. In each case a complicated combination of
alternating sums collapses to MZVs, using the datamine.
  
\subsection{7 loops:  Erik Panzer's polylogs of the sixth root of unity}

For the period $P_{7,11}$ the Stembridge count $c(p)$ takes different
forms according as the residue of $p$ mod 3~\cite{S2} as may be
quickly determined using the $c_2$ invariant of~\cite{BSY}. So we expect
a reduction to weight-11 multiple polylogs of the 6th root of unity. There are 
$6^2\times7^9=1,452,729,852$  of these with legal words.

Erik Panzer recently reported~\cite{P} a reduction of $P_{7,11}$
to a subset of these words and hence obtained several thousand digits
of $P_{7,11}$, completing the 7-loop challenge. 
We have checked
this answer to 21 digits, by intensive numerical methods. 
A compact presentation of Panzer's lengthy result
has been obtained by Schnetz, using algebraic methods based on a
co-action for polylogarithms. Moreover, Panzer was able to use PSLQ
to reduce his result to a basis of dimension 72,
resulting from applying the generalized parity conjecture of~\cite{B1}
to a three-letter Deligne alphabet.

\subsection{8 loops: planar $N=4$ supersymmetric Yang-Mills theory}

Without evaluating Feynman diagrams, 
S\'ebastien Leurent and Dmytro Volin have computed the Konishi anomalous dimension
of planar $N=4$ SYM up to 8 loops~\cite{LV}. An MZV first appears at 8 loops,
where the weight-11 term 
$$\frac{864g^{16}}{5}\left\{76307\zeta(11)+792[\zeta(3,5,3)-\zeta(3)\zeta(5,3)]
-18840\zeta(3)^2\zeta(5)\right\}$$
is happily reducible to the 3 terms found in $\phi^4$ theory
almost 20 years ago~\cite{BK1,BK2}.  

{\bf Open question}: Do MZVs suffice for the next ``wrapping'' at 12 loops? 

\subsection{8 loops: further reductions to  polylogs}

There are 41 targets in the census at 8 loops~\cite{S1}.
Thanks to recent progress with graphical functions~\cite{S3} 31 of these have been evaluated
in terms of polylogs. All periods for which weight-drop was predicted have been reduced to
MZVs of weight less than 13.
One period, $P_{8,16}$, exhibits
\emph{weight mixing}, evaluating to a
mixture of weight-10 and weight-11 MZVs.  Schnetz has evaluated the period
$P_{8,33}$ as a compact combination of weight-13 polylogs of the 6th root of unity.
Of the 10 periods so far unidentified at 8 loops,
it is anticipated  that 6 may eventually be evaluated in terms of MZVs
and polylogs of 4th or 6th roots of unity.  
  
\subsection{8, 9 and 10 loops: modular obstructions to polylogs}

Brown and 	Schnetz~\cite{BS2} have identified many periods with more than 7 loops
for which there is strong evidence that polylogs cannot suffice
for evaluations. There is no chain of integration over Schwinger parameters
that satisfies the criterion of ``linear independence" explained in~\cite{P}.
In 16 cases, point counts in finite fields of the non-linear denominators
correspond, modulo primes, to the Fourier coefficients of well known modular forms. 
In particular
$(\eta_1\eta_7)^3$, $(\eta_1\eta_5)^4$, 
$(\eta_1\eta_8)^2\eta_2\eta_4$ and $(\eta_1\eta_3)^6$
are the modular forms that prevent reduction to polylogs
of 4 subdivergence-free diagrams that give scheme-independent
contributions to the 8-loop beta-function of $\phi^4$ theory.
For these, one may need a theory  of ``multiple modular values"
whose details are at present obscure. 

{\bf Acknowledgements:} We thank our colleagues Spencer Bloch, Francis Brown,
Dzmitry Doryn, Dirk Kreimer, Erik Panzer and Karen Yeats for many
illuminating discussions of the periods of quantum field theory
and their relationship to algebraic geometry.


\begin{thebibliography}{99}

\bibitem{BBBG}
D.H.~Bailey, J. M.~Borwein, D.~Broadhurst, M. L.~Glasser,
\emph{Elliptic Integral Evaluations of Bessel Moments},
\emph{J. Phys.} {\bf A41} (2008) 205203 
[{\tt arXiv:0801.0891}].

\bibitem{BV}
S.~Bloch, P.~Vanhove,
\emph{The elliptic dilogarithm for the sunset graph}
[{\tt arXiv:1309.5865}].
  
\bibitem{BKV}
S.~Bloch, M.~Kerr, P.~Vanhove,
\emph{A Feynman integral via higher normal functions}
[{\tt arXiv:1406.2664}].

\bibitem{BBV}
J. Bl\"umlein, D. J. Broadhurst, J. A. M. Vermaseren,
\emph{The Multiple Zeta Value data mine},
\emph{Comput. Phys. Commun.} {\bf 181} (2010) 582-625,
[{\tt arXiv:0907.2557}].

\bibitem{B1}
D.J.~Broadhurst,
\emph{Massive 3-loop Feynman diagrams reducible to SC* primitives of algebras of the sixth root of unity},
\emph{Eur. Phys. J.} {\bf C8} (1999) 311-333 
[{\tt  arXiv:hep-th/9803091}].

\bibitem{B2}
D.~Broadhurst,
\emph{Elliptic integral evaluation of a Bessel moment by contour integration of a lattice Green function}
[{\tt arXiv:0801.4813}].

\bibitem{B3}
D.~Broadhurst,
\emph{Multiple Zeta Values and Modular Forms in Quantum Field Theory}
in \emph{Computer Algebra in Quantum Field Theory} 33-72, edited by
C.~Schneider and J.~Bl\"umlein (Springer) 2013.

\bibitem{BFT}
D.J.~Broadhurst, J.~Fleischer, O.V.~Tarasov,
\emph{Two-loop two-point functions with masses: asymptotic expansions and Taylor series, in any dimension},
\emph{Z. Phys.} {\bf C60} (1993) 287-302
[{\tt arXiv:hep-ph/9304303}].

\bibitem{BK1}
 D.J.~Broadhurst, D.~Kreimer,
\emph{Knots and numbers in $\phi^4$ theory to 7 loops and beyond},
\emph{Int. J. Mod. Phys.} {\bf C6} (1995) 519-524
[{\tt arXiv:hep-ph/9504352}].

\bibitem{BK2}
D.J.~Broadhurst, D.~Kreimer,
\emph{Association of multiple zeta values with positive knots via Feynman diagrams up to 9 loops},
\emph {Phys. Lett.} {\bf B393} (1997) 403-412
[{\tt arXiv:hep-th/9609128}].
 
\bibitem{BS1}
F.~Brown, O.~Schnetz,
\emph{A K3 in $\phi^4$},
\emph{Duke Math. J.} {\bf 161} (2012) 1817-1862
[{\tt arXiv:1006.4064}].

\bibitem{BS2}
F.~Brown, O.~Schnetz,
\emph{Modular forms in Quantum Field Theory},
\emph{Commun. Number Theory Phys.} {\bf 7} (2013) 293-325
[{\tt arXiv:1304.5342}].

\bibitem{BSY}
F.~Brown, O.~Schnetz, K.~Yeats,
\emph{Properties of $c_2$ invariants of Feynman graphs}
[{\tt arXiv:1203.0188}]

\bibitem{DLST}
F.~Di Renzo, M.~Laine, Y.~Schr\"oder, C.~Torrero,
\emph{Four-loop lattice-regularized vacuum energy density of the three-dimensional $SU(3)$ + adjoint Higgs theory},
\emph{JHEP} {\bf  9} (2008) 061
[{\tt arXiv:0808.0557}].

\bibitem{D}
D.~Doryn,
\emph{On one example and one counterexample in counting rational points on graph hypersurfaces},
 \emph{Lett. Math. Phys.} {\bf 97} (2011) 303-315
[{\tt arXiv:1006.3533}].

\bibitem{LV}
S.~Leurent, D.~Volin,
\emph{Multiple zeta functions and double wrapping in planar $N=4$ SYM},
\emph{Nuclear Physics} {\bf B875}  (2013)  757-789
[{\tt  arXiv:1302.1135}].

\bibitem{P}
E.~Panzer,
\emph{Feynman integrals via hyperlogarithms}
[{\tt arXiv:1407.0074}].

\bibitem{S1}
O.~Schnetz,
\emph{Quantum periods: A census of $\phi^4$-transcendentals},
\emph{Commun. Number Theory Phys.} {\bf 4} (2010) 1-48
[{\tt arXiv:0801.2856}].

\bibitem{S2}
O.~Schnetz,
\emph{Quantum field theory over ${\bf F}_q$},
\emph{Electron. J. Combin.} {\bf 18} (2011) 102
[{\tt arXiv:0909.0905}].

\bibitem{S3}
O.~Schnetz,
\emph{Graphical functions and single-valued multiple polylogarithms}
[{\tt arXiv:1302.6445}].

\bibitem{St}
J.~Stembridge,
\emph{Counting points on varieties over finite fields related to a conjecture of Kontsevich},
\emph{Ann. Combin.} {\bf 2} (1998) 365-385.

\end{thebibliography}
\end{document}